\documentclass[prl,twocolumn,superscriptaddress]{revtex4}
\usepackage{graphicx}
\usepackage{amsmath,amssymb}
\usepackage{hyperref}
\usepackage{amsmath}
\usepackage{amsfonts}
\usepackage{amssymb}
\providecommand{\U}[1]{\protect\rule{.1in}{.1in}}
\ifx\pdfoutput\relax\let\pdfoutput=\undefined\fi
\newcount\msipdfoutput
\ifx\pdfoutput\undefined\else
\ifcase\pdfoutput\else
\ifx\paperwidth\undefined\else
\ifdim\paperheight=0pt\relax\else\pdfpageheight\paperheight\fi
\ifdim\paperwidth=0pt\relax\else\pdfpagewidth\paperwidth\fi
\fi\fi\fi
\begin{document}
\title{Observation of blue-shifted ultralong-range Cs$_{2}$ Rydberg molecules}
\author{J. Tallant}
\affiliation{Homer L. Dodge Department of Physics and Astronomy, The University of
Oklahoma, 440 W. Brooks St. Norman, OK 73019, USA}
\author{S.~T.~Rittenhouse}
\affiliation{ITAMP, Harvard-Smithsonian Center for Astrophysics, 60 Garden Street,
Cambridge, Massachusetts 02138}
\author{D. Booth}
\affiliation{Homer L. Dodge Department of Physics and Astronomy, The University of
Oklahoma, 440 W. Brooks St. Norman, OK 73019, USA}
\author{H.~R.~Sadeghpour}
\email{[]hsadeghpour@cfa.harvard.edu}
\affiliation{ITAMP, Harvard-Smithsonian Center for Astrophysics, 60 Garden Street,
Cambridge, Massachusetts 02138}
\author{J. P. Shaffer}
\email{[]shaffer@nhn.ou.edu}
\affiliation{Homer L. Dodge Department of Physics and Astronomy, The University of
Oklahoma, 440 W. Brooks St. Norman, OK 73019, USA}
\date{\today}
\begin{abstract}
We observe ultralong-range blue-shifted Cs$_{2}$ molecular states near $ns_{1/2}$ Rydberg states
in an optical dipole trap, where $31\leq n\leq34$. The accidental near degeneracy of $(n-4)l$ and $ns$ Rydberg states for $l>2$ in Cs, due to the small fractional $ns$ quantum defect, leads to non-adiabatic coupling among these states, producing potential wells above the $ns$ thresholds. Two important consequences of admixing high angular momentum states with $ns$ states are the formation of large permanent dipole moments, $\sim 15-100\,$Debye, and accessibility of these states via two-photon association. The observed states are in excellent agreement with theory. Both projections of the total angular momentum on the internuclear axis are visible in the experiment.
\end{abstract}
\maketitle

The observation of ultralong-range Rydberg molecules has piqued interest in few-body Rydberg
interactions that occur in ultracold atomic gases \cite{Bend09,Over09,Bend10}. The interaction for an exotic class of these systems, trilobite molecules, arises to first order from the zero-range pseudopotential scattering of a Rydberg electron from a ground state perturber. Generally, bound molecular states form when the
electron-atom scattering length is negative. Part of the fascination with such molecules originates from the
prediction that these molecules can possess massive permanent electric dipole
moments, $\sim1\,$kDebye, making them amenable to electric field manipulation
\cite{Greene00}. Large dipole moments are expected when high electron angular momentum
hydrogenic degenerate manifolds are involved in the electron-atom scattering. Ultracold Rydberg molecules formed by two-photon
association into Rb($ns$) states were expected only to have an induced dipole
moment \cite{Bend09}. It was later demonstrated theoretically
and experimentally \cite{Li11} that due to the near integer quantum defect
of Rb($ns$), $\mu_s = 3.13$, nearby and nearly degenerate $(n-3)$ manifolds mixed highly localized trilobite-like states into the zeroth order symmetric $ns$ states, giving them an observable permanent dipole moment, $\sim1\,$Debye,
in the body-fixed frame.

\begin{figure}[ptbh]
\begin{center}
\includegraphics[height=2.25in]{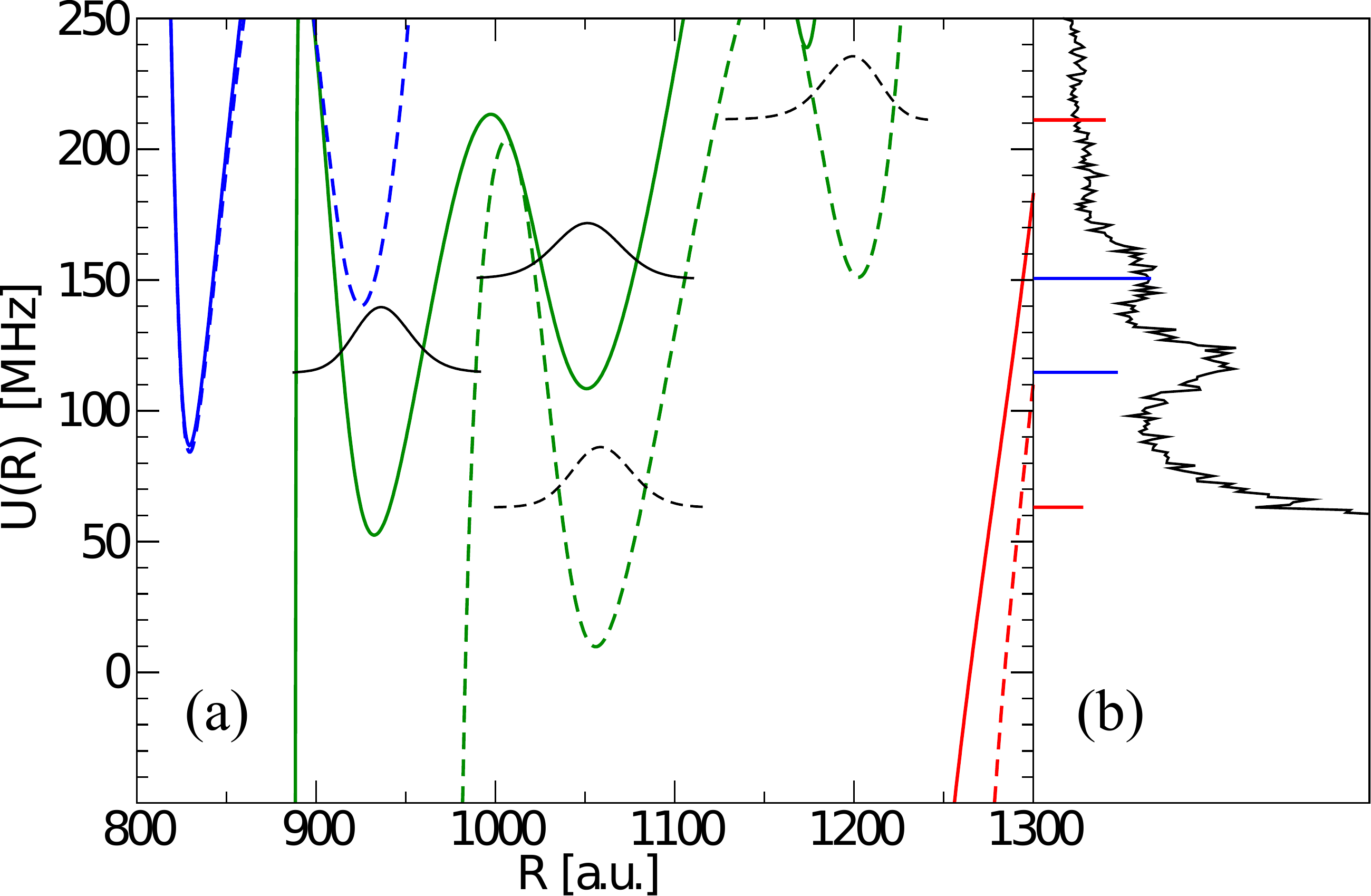}
\caption{(Color Online) (a) The $n=31$ BO potential energy curves for  $M_{J}=\pm
1$ (solid curves) and $M_J=0$ (dashed curves) Cs$_2$ Rydberg molecules in the
vicinity of the $ns$ Rydberg threshold. Also shown are the low lying
vibrational wave functions in the experimentally accessible region. The zero of energy is set to the appropriate
$31s_{1/2}$ atomic Rydberg energy. (b) The experimentally observed spectra are shown
in arbitrary units compared to the BO potentials. The line sticks give the positions of the predicted energies and relative transition strengths of the
predissociating vibrational states. The two $M_J=\pm 1$ vibrational levels have permanent dipole moments of 33.5 and 37.4 D, respectively. }
\label{fig:31spec}
\end{center}
\end{figure}
Here, we report the spectroscopic observation of ultralong-range trilobite-like Rydberg molecules formed in an ultracold gas of Cs with large permanent electric dipole moments. The spectroscopic signature of these molecules is in excellent agreement with theoretical predictions and serves as a milepost for creating trilobite molecules with giant dipole moments. The unique features of these ultralong-range Rydberg molecules are threefold. First, these states are unexpectedly blue-shifted with respect to the Cs($ns$) thresholds. This feature results from the fact that the fractional part of the Cs$(ns$) quantum defect is small,  $\mu_s = 4.05$,  and $(n-4)l$, $l>2$, ``trilobite" manifolds become nearly degenerate, and mix strongly with the $ns$ states \cite{Li11}, inducing non-adiabatic avoided crossings of the $ns$-dominated Born-Oppenheimer (BO) potential energy curves. Second, the energetic proximity to the degenerate high-angular momentum hydrogenic states results in the molecular wavefunction having hydrogenic character at the $1\%$ level. Therefore, large permanent electric dipole moments ($15-100$ Debye) are predicted for these molecules. The rather large hydrogenic character of the wave function is promising for generating trilobite molecules with kD dipole moments. Third, the position
of the Cs trilobite states depends sensitively on the location of the $p$-wave $e^{-}$-Cs scattering resonance \cite{Khusk02}, and therefore the present spectra also serve as an indirect probe of the accuracy of the $^{3}P_{1}$ resonance position \cite{Scheer97}. Unlike atoms in magnetic traps, both parallel and perpendicular projections of the magnetic quantum number (i. e. $M_J=0,\pm 1$) are visible in this experiment because it is carried out in an optical dipole trap. These molecules dissociate to their parent atomic constituents.

In Fig. \ref{fig:31spec}, we present the main result of this work. The observed blue-shifted Cs Rydberg molecular spectrum accurately correlates with the vibrational lines of the calculations for states near the Cs$(31s)$ threshold. Both $M_J=0 ,\pm 1$ molecular symmetries are present. The blue-shifted potential wells arise from the strong non-adiabatic BO coupling of the $31s$ state with the localized $27l$, $l>2$, states. The binding energies are most easily defined with respect to the $ns$ threshold as each BO potential is largely $s$-character in the region where the spectroscopic resonances are observed. The molecular features are broadened with increasing energy from the $31s$ threshold. This can be understood to be the result of progressively larger permanent dipole moments, as the higher energy states have more high $l$ character in their electronic wave functions.

Relatively large densities and ultracold temperatures are required to efficiently create the ultralong-range Cs molecules. The Cs molecules are photoassociated inside a crossed far off-resonance
trap (FORT) which is loaded from a Cs vapor cell MOT. The experimental setup
has been described elsewhere \cite{Tall10}. The crossed FORT is prepared from
a 10 W 1064 nm Yb fiber laser. The first arm of the cross is formed by
copropagating the FORT beam with one axis of the MOT trapping laser. The Yb
laser beam is focused to a spot size ($1/e^{2}$) of 86 $\mu$m $\pm\,1.1$ $\mu
$m. The focal spot size was verified with a CCD camera and the results in
Ref.~\cite{Tall10}. The first Yb beam is recycled and focused back into the chamber
with the same spot size. The Yb beam intersects the first beam of the FORT at
22.5$^{\circ}$ creating a crossed trapping region which is slightly cigar
shaped with a 2:1 aspect ratio. At 17 W of power, the FORT depth is $\sim1.5$ mK. The highest trap frequency
is $2\pi\cdot2.23$ kHz and the lowest trap frequency is $2\pi\cdot445$ Hz. The crossed FORT is loaded as in \cite{Tall10} to a peak density of $\sim
2\times10^{13}$ cm$^{-3}$. The density was verified with absorption imaging
and Rydberg atom detection rates. To account for the combined ac Stark shifts of the ground and
Rydberg states, the position of the Rydberg state was measured inside the MOT
and compared to the position of the Rydberg state in the crossed FORT. The
measured average ac Stark shift is 19 MHz.

The molecular states are
excited using a two-photon process. The first step of the excitation is an
infrared (IR) photon which is tuned 182$\,$MHz red of the Cs $6p_{3/2}$ hyperfine
manifold. The second photon is generated by a ring dye laser tuned near 512$\,$nm. The infrared beam is sent through an acousto-optic modulator (AOM) and single-mode
polarization-preserving fiber. The output is collimated to a size of 1$\,$mm$^{2}$ and intersects the crossed FORT at an angle of 79$^{\circ}$ with
respect to the long axis of the cross. During excitation, this beam has 5$\,$mW
of power. The green beam is also sent through an AOM and a single-mode
polarization-preserving fiber. The output is copropagated with the second FORT
beam and is focused onto the crossed region with a spot size of 44$\,\mu$m. The
power used for excitation is 70$\,$mW. The two-photon linewidth of the excitation
pulses was measured inside of the MOT to be $<3\,$MHz.


To locate the positions of the molecular states, the green laser is scanned on
the blue side of the $ns_{1/2}$ Rydberg states while the IR laser remains
locked $182\,$MHz below the Cs $6s_{1/2}(F=3)\rightarrow6p_{3/2}(F=2)$
transition. An absorption spectrum is acquired by monitoring the number of
ions produced as a function of green laser frequency. The excitation pulses
begin 20 ms after the crossed FORT has been loaded to let the uncaptured MOT
atoms fall away. Each excitation pulse is 10$\,\mu$s long and is immediately
followed by an electric field pulse to project any positive ions onto a
microchannel plate detector where they are counted. The excitation step
repeats at $1.0\,$kHz and lasts 500$\,$ms, at which time the green laser frequency
is incremented by 1$\,$MHz and the crossed FORT is reloaded. The electric field
used to extract the ions, 67$\,$V$\,$cm$^{-1}$, is far below the ionization
threshold of any Rydberg states in the experiment \cite{Gallagher}. The FORT
beam is used to photoionize any Rydberg atoms as well as the Rydberg atom
constituents of the ultralong-range molecules. Ion signals corresponding to
the Cs$_{2}^{+}$ molecular time-of-flight are simultaneously acquired with the
Cs$^{+}$ signal. As many as six absorption spectra are averaged together with scaled
frequencies to obtain a single experimental spectrum. Ions arriving at the
molecular time-of-flight were not observed, suggesting a different decay
mechanism than found in Ref.~\cite{Bend09}. However, if the molecules gain $>1\,$GHz
of energy, they will be lost from the ionization region.

Precise knowledge of the absolute frequencies of both excitation lasers is
required to correctly describe the energies of observed molecular transitions.
The absolute frequency of the IR beam is monitored with a saturated absorption
setup. The light for the saturated absorption is shifted with AOMs such that the Cs $6s_{1/2} (F=3) \rightarrow6p_{3/2} (F=3)$
transition is on resonance during the experiment. The saturated absorption
spectrum determines the IR frequency within $\sim3$ MHz. To monitor the
frequency of the green laser, a fraction of the laser output is combined with
light from the IR laser to generate an electromagnetically induced
transparency (EIT) signal in a room-temperature Cs vapor cell. The EIT setup
is arranged in both co- and counterpropagating configurations with respect to
the IR beam. This arrangement produces up to six EIT resonances corresponding
to different hyperfine states and detunings whose absolute frequency relations
are known. The positions of the EIT resonances can be shifted to any desired
location with AOMs to provide spectroscopic markers.


The bonding of these types of Rydberg molecules derives from the frequent scattering of the
highly excited Rydberg electron from the ground state atom perturber
\cite{Greene00,Bah01}. The relevant electron-ground state atom interaction is
described within the zero-range Fermi pseudopotential approximation
\cite{Fermi36,Omont77}, given in atomic units by
\begin{eqnarray}
V_{e-a}\left(  \mathbf{r}\right)  =2\pi A_{s}\left(  k\right)  \delta^{\left(
3\right)  }\left(  \mathbf{r}-\mathbf{R}\right) + \label{Eq:eaint}\\ \nonumber 6\pi A_{p}^{3}\left(
k\right)  \delta^{\left(  3\right)  }\left(  \mathbf{r}-\mathbf{R}\right)
\overleftarrow{\nabla}\cdot\overrightarrow{\nabla}
\end{eqnarray}
where $A_{s}$ and $A_{p}$ are the energy dependent $s$- and $p$-wave
scattering length and $\mathbf{r}$ and $\mathbf{R}$ denote the
positions of the electron and ground state atom relative to the Rydberg core,
respectively. The BO potentials which support the molecular predissociating
states are found by diagonalizing the interaction of Eq. (\ref{Eq:eaint}) in a
basis of Rydberg electron $nl$ orbitals \cite{Khusk02,Li11}. Within the semi-classical picture the electron wavenumber $k$ is
related to the intermolecular distance $R$ in atomic units by $k=\sqrt
{2\left(  1/R-E_{b}\right)  }$ where $E_{b}$ is the binding energy of the
isolated Rydberg atom. The energy dependent scattering lengths are given by
$A_{s}\left(  k\right)  =-\tan\delta_{s}/k$ and $A_{p}^{3}\left(  k\right)
=-\tan\delta_{p}/k^{3}$ where $\delta_{l}$ is the scattering phase shift for
$l$-wave electron Cs scattering. Cs has relatively large spin-orbit interaction resulting in different ($J=0,1$ and 2) $e^{-}$-Cs $^{3}P_{J}$ resonant scattering phase shifts. To account for this, we take the $p$-wave
scattering to be,
\begin{equation}
A_{p}^{3}=\sum_{J=0}^{2}\left[  C_{10,1M_{J}}^{JM_{J}}\right]  ^{2}A_{p,J}^{3}
\end{equation}
where $C_{L_{1}M_{1},L_{2}M_{2}}^{JM_{J}}$ is a Clebsch-Gordan coefficient coupling the bound $e^-$ angular momenta ($L_1M_1$) to the scattering $e^-$ angular momenta ($L_2M_2$) to form the total angular momenta ($JM$). The
different Clebsch-Gordan coefficients create different potentials for
the $M_{J}=\pm1$ and 0 cases. The $p$-wave scattering lengths
are found by setting the $^{3}P_{1}$ resonance position to $8$ meV
\cite{Scheer97} and the splitting of the $^{3}P_{0}$ and $^{3}P_{2}$ states to
those of Ref. \cite{Thumm91}.

\begin{figure*}[ptb]
\includegraphics[width=6.5in]{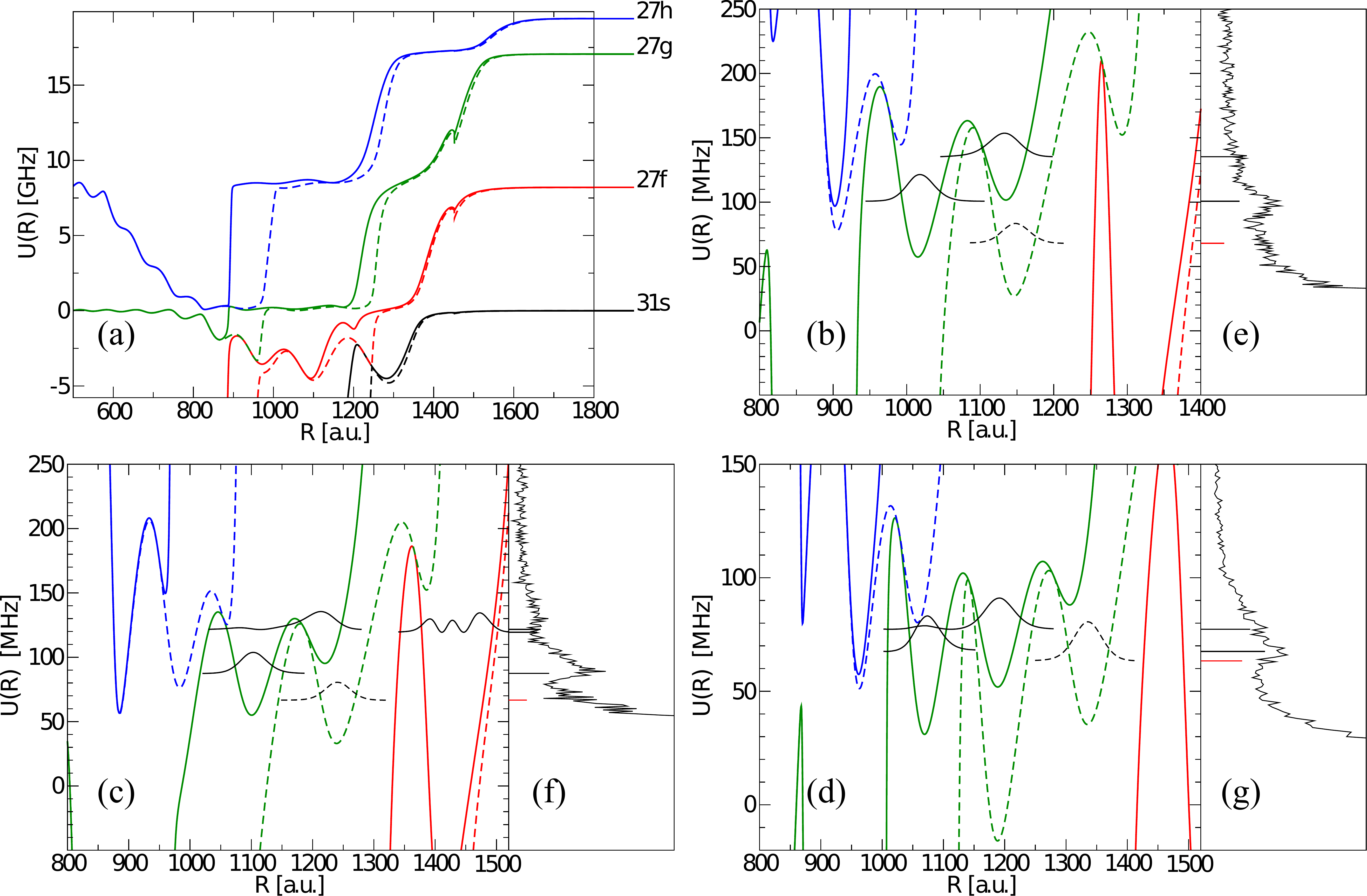}
\caption{(Color Online) (a) The BO potential energy curves for $M_{J}=\pm
1$ (solid curves) and $M_J=0$ (dashed curves) correlating to atomic Rydberg thresholds in the vicinity of Cs($31s$) state. The $n=32-34$ BO potential energy curves for  $M_{J}=\pm
1$ (solid curves) and $M_J=0$ (dashed curves) are displayed in (b-d) and the corresponding observed spectra are shown in (e-g), respectively. For detailed description of the features, see caption for Fig.~\ref{fig:31spec}.  The zero on the energy axis refers to the corresponding Cs($ns$) threshold.
}
\label{fig:resultspanel}
\begin{center}
\end{center}
\end{figure*}

The resulting BO potentials near the $ns$ Rydberg threshold are shown for
$n=31-34$ in Figs.~\ref{fig:31spec} and \ref{fig:resultspanel}(b-d) respectively for total angular
momentum projection $M_{J}=\pm1$ (solid curves) and 0
(dashed curves). Fig.~\ref{fig:resultspanel}(a) shows the $n=31$ BO potentials over a larger energy range correlating to atomic Rydberg thresholds in the vicinity of Cs($31s$) state.
The near degeneracy of the $(n-4) l>2$ manifolds with the $ns$ series ensures that there are non-adiabatic couplings between the molecular potential energy curves of the same symmetry. Two important consequences of this interaction are the formation of potential wells capable of supporting bound states, blue-shifted with respect to the $ns$ Rydberg thresholds, and accessibility of these states via two-photon association. The latter is facilitated by the fact that the electronic wave functions for the molecular states contain large $s$-wave components.
Figs.~\ref{fig:resultspanel}(b-d) also show several of the low lying vibrational
wavefunctions within the experimentally accessible region. The averaged experimental data
near the $ns_{1/2}$ Rydberg states for $n=31-34$ are compared with the
theoretical predictions and show excellent agreement. All of the molecular states have spatial angular momentum projection $M_{L}=0$ \cite{Greene00,Khusk02}, i.e. all of the
states are $^{3}\Sigma$ molecules. While there exist BO potential energy curves
corresponding to the $^{3}\Pi$ symmetry, they do not couple to the molecular
states considered here \cite{Khusk02}. Because the BO potentials near
the $ns$ thresholds are sensitive to the position of the $^{3}P_{1}$
resonance, the agreement between theory and experiment can be used as an indirect probe of the accuracy of the resonance position \cite{Scheer97}. Most of the observed states correspond to
the $M_{J}=\pm1$ projections of the electronic angular momentum, but because
these states are photoassociated in a FORT, we are sensitive to the $M_{J}=0$
projection as well. In all the spectra, we observe features corresponding to $M_{J}=0$ states.

The molecules also experience an ac Stark shift in the FORT. As the
molecular states consist of single Rydberg excitations, and the ac Stark shift
of a Rydberg atom is only weakly dependent on the Rydberg state
\cite{Gallagher,Younge10}, we then expect that the Rydberg molecular states
have the same overall shift and inhomogeneous broadening seen in the
bare atomic transition. This hypothesis is corroborated by the fact that the observed data shifted by the measured atomic ac Stark shift, $19\,$MHz, are in good agreement with theory. However, the experimental widths of the most prominent states seen in Figs.~\ref{fig:resultspanel}(e-g) are $\sim20$ MHz, far larger than that expected from the atomic Rydberg linewidth, $\sim100$ kHz \cite{But10} and that observed for the Cs Rydberg atoms, $\sim 11\,$MHz. The additional broadening is likely due to the existence of large dipole moments for the Rydberg molecules, $\sim30$ D.
We expect small stray electric fields (on the order of 100$\,$mV$\,$cm$^{-1}$) coupled to the large molecular dipoles to produce broadening in the range of $4-10\,$MHz.
The strength of the dipole moments increases with increasing energy above the Cs($ns$) thresholds, as the molecular states progressively retain less and less $s$-wave character. The observed molecular spectra in Fig.~\ref{fig:31spec}(b) and \ref{fig:resultspanel}(e-g) become progressively broadened with energy, lending further support to our assertion.


A molecule with a dipole moment of $\sim 35\,$D should exhibit a dramatic linear
Stark effect. A modest applied field of 240 mV cm$^{-1}$ can produce a
measurable shift. To investigate this, small background
electric fields were applied during excitation of the two $M_{J}=\pm1$ molecular states near $n=31$. Surprisingly, even electric
fields below 100 mV cm$^{-1}$ cause changes in the observed spectrum. A linear Stark effect is not observed. Rather, the
observed peaks broaden and slightly shift to the blue ($\sim10$ MHz) before
the bluest peak disappears completely. The additional broadening of the
lines decrease their amplitude and complicate the measurement. The effect is
different for each of the observed molecular states, which is to be expected
based on their different potential minima. Our observations cannot be simply
described as a linear Stark shift because the potentials supporting the bound
states are changing appreciably with the applied field. This is expected to be
different than the Rb case \cite{Li11} because the molecular states here have
significant mixing with the degenerate hydrogenic manifold and are the result
of avoided crossings, which are sensitive to the applied electric field. As a consequence, the electric field dependence of
the Cs ultralong-range molecules studied is sufficiently complicated to prevent
a full analysis in this paper.

In summary, we have observed Cs Rydberg molecular states in a crossed optical
dipole trap and found the spectrum to be in excellent agreement with theory.
The agreement with theory provides an experimental probe for the position of
the $p$-wave resonance \cite{Scheer97}. We observed both
$M_{J}=0$ and $M_{J}=\pm1$ states, not easily observed in a magnetic trap. The
molecular states studied have different properties than those reported earlier
for Rb. There is significantly large mixing of the
highly localized \textquotedblleft trilobite-like\textquotedblright\ electronic character in Cs.
The additional line broadening observed is believed to arise from the
interaction of the large permanent electric dipole moment.
We
measured large changes in the spectrum by applying small electric fields,
suggestive of both large permanent electric dipole moments and structural
changes to the states, consistent with theory. In contrast with Rb, a large signal for the Cs$_{2}^{+}$ molecular ion is not observed, suggesting
a different decay mechanism. We will attempt to address  the behavior of the molecular states in electric fields and understand
their decay mechanisms. The discovery of spectroscopically accessible Cs
trilobite states opens another window into these exotic molecules. It may be possible to use these states to create ion pair states. Applications that exploit their large permanent dipole moments are sure to follow.

\section*{Acknowledgements}
The authors thank C.~H.~Greene, T. Pfau, and L. Marcassa for helpful discussions. HRS and STR acknowledge support through an NSF grant to ITAMP at Harvard University and the Smithsonian Astrophysical Observatory. The experiment was supported by the NSF (PHY-0855324).

\bibliographystyle{apsrev4-1}

\end{document}